\documentclass[preprint,aps]{revtex4}

\begin{document}

\title{2D Calogero Model in the Collective-Field Approach}
\author{Velimir Bardek}
\email{bardek@irb.hr}
\author{Danijel Jurman}
\email{djurman@irb.hr}
\affiliation{Theoretical Physics Division,\\
Rudjer Bo\v skovi\'c Institute, P.O. Box 180,\\
HR-10002 Zagreb, CROATIA}

\begin{abstract}
We consider the large-N Calogero-Marchioro model in two 
dimensions in the Hamiltonian collective field approach based on
the $1/N$ expansion. The Bogomol'nyi limit appears in the 
presence of the harmonic confinement. We investigate density
fluctuations around the semiclassical uniform solution.
The excitation spectrum splits into two branches depending on the 
value of the coupling constant. The ground state exhibits long-range
order.
\end{abstract}

\maketitle

\section{Introduction}
The generalized quantum N-body Calogero-Marchioro (CM) model in two and
 arbitrary number
of dimensions has been in focus of research recently 
\cite{Calogero2,Khare1,Basu,Murthy}.
There are many exact solutions and their properties known by now.
However, the analytical solution of the full many-body CM problem in two
dimensions is not tractable.
The quantum problem may not be solvable even for the ground state
\cite{Date1} for arbitrary values of coupling strenghts of the two
and three-body interactions.
 Thow it is possible to find an infinite number of
exact radial eigenstates, the set is not complete 
\cite{Khare2,Bhaduri,Ghosh2,Khare3,Ghosh3,Date2}.
The two-dimensional CM model deserves special attention due to its
relation to quantum dots, quantum Hall effect, random matrix theory,
anyon physics and other systems of physical interest \cite{Feigel,Oas,Kane}. 
Apart of its physical application, we believe that CM model
in two dimensions deserves a careful field theoretical analysis.
 In this paper we investigate the large-N CM model in the
bosonic picture using the large-N collective field technique.

\section{Collective field Hamiltonian}
The 2D Calogero Hamiltonian describes a system of N non-relativistic
particles on a plane, interacting via the two-body inverse-square potential
and a long-range three-body interaction. It is given by
\begin{equation}
H=-\frac{1}{2}\sum_{i} \vec{\nabla}_{i}^{2}
 + g \sum_{i\neq j}\frac{1}{|{\bf r}_{i}-{\bf r}_{j}|^{2}} + 
G \sum_{i \neq j \neq k} \frac{ ({\bf r}_{i}-{\bf r}_{j})
({\bf r}_{i}-{\bf r}_{k})}
{|{\bf r}_{i}-{\bf r}_{j}|^{2} |{\bf r}_{i}-{\bf r}_{k}|^{2}}
+ \frac{{\omega}^{2}}{2} \sum_{i=1} {{\bf r}_{i}}^{2}\;\;.
\end{equation}
The particles are confined in a one-body oscillator potential. We 
 choose units such that $\hbar=m=1$; g and G are in general arbitrary  
positive dimensionless coupling strenghts of the two and the three-body
interactions.\\
\indent
The singularity of the Hamiltonian (1) at
 ${\bf r}_{i}={\bf r}_{j}$ requires the wave function to have a prefactor
of Jastrow-type which vanishes for coincident particles.
We extract this prefactor in the form
\begin{equation}
\psi({\bf r}_{1},{\bf r}_{2},...,{\bf r}_{N})=
\prod_{i<j}{|{\bf r}_{i}-{\bf r}_{j}|}^{\lambda}
\phi({\bf r}_{1},{\bf r}_{2},...,{\bf r}_{N})
\end{equation}
and obtain the new Hamiltonian
\[
H=-\frac{1}{2}\sum_{i} \vec{\nabla}_{i}^{2}
 -\lambda \sum_{i\neq j}
\frac{{\bf r}_{i}-{\bf r}_{j}}{|{\bf r}_{i}-{\bf r}_{j}|^{2}}\vec{\nabla}_{i}+
\left(g-\frac{\lambda^{2}}{2}\right)
\sum_{i\neq j}\frac{1}{|{\bf r}_{i}-{\bf r}_{j}|^{2}} +
\] 
\begin{equation}
+\left(G-\frac{\lambda^{2}}{2}\right)
 \sum_{i \neq j \neq k} \frac{ ({\bf r}_{i}-{\bf r}_{j})
({\bf r}_{i}-{\bf r}_{k})}
{|{\bf r}_{i}-{\bf r}_{j}|^{2} |{\bf r}_{i}-{\bf r}_{k}|^{2}}
+ \frac{{\omega}^{2}}{2} \sum_{i} {{\bf r}_{i}}^{2}\;\;,
\end{equation}
acting on the residual wave function $\phi$.
The above Hamiltonian (3) simplifies drastically in the limit
$g=G=\lambda^2/2$ which we use from now on.
This is precisely the case when the quantum mechanical ground state
and an infinite tower of some excited eigenstates may be obtained exactly
\cite{Calogero2}.
We are now in a position to introduce the collective field and to formulate
a non-relativistic quantum field theory which operates in a large N-particle
sector of the Hilbert space \cite{Jevicki}.
 The collective field is assumed to be the 
density function
\begin{equation}
\rho({\bf r})=\sum_{i} \delta({\bf r}-{\bf r}_{i})\;\;,
\end{equation}
where ${\bf r}_{i}$'s are positions of $N$ spinless bosonic particles.
From the definition (4) it follows that the collective field obeys the 
normalization condition
\begin{equation}
\int d^{2}r \rho({\bf r})=N\;\;.
\end{equation}
 
Next, we reformulate the differential operators $\vec{\nabla}_{i}$  
in the Hamiltonian (3) in terms of a functional differentiation with
respect to the collective field $\rho({\bf r})$
\begin{equation}
\vec{\nabla}_{i}=\int d^{2}r \left(\vec{\nabla}_{i}\rho({\bf r})\right)
\frac{\delta}{\delta\rho({\bf r})}\;\;.
\end{equation}
In the large-N limit, the Hamiltonian (3) can be expressed entirely in terms
of the collective field $\rho({\bf r})$ and its canonical conjugate
\begin{equation}
\pi({\bf r})=-i\frac{\delta}{\delta\rho({\bf r})}\;\;.
\end{equation}
The continuum limit of the Hamiltonian (3) can be written as
\[ 
H=\frac{1}{2}\int d^2r \rho({\bf r}) (\vec{\nabla}\pi)^2+
\frac{\omega^{2}}{2} \int d^2 r \rho({\bf r}) {\bf r}^{2}+
\]
\begin{equation}
+i\lambda \int d^2r \left(\vec{\nabla}\int d^2 r' \rho({\bf r})
\frac{{\bf r}-{\bf r'}}{|{\bf r}-{\bf r'}|^{2}} 
 \rho({\bf r'})\right) \pi({\bf r})+
i\frac{\lambda-2}{4}\int d^2 r \left(\Delta \rho({\bf r})\right) \pi({\bf r})
\;\;.
\end{equation}
In order to re-express all the sums over
 $i$ and $j$ in the Hamiltonian (3) in terms of the collective field
 $\rho({\bf r})$, we have been forced to include terms $i=j$, too. These terms,
however, are superfluous in the correct form of the Hamiltonian and therefore 
should be substracted. This is the origin of the  $\lambda$
 dependent part of the 
last term in the Hamiltonian (8) \cite{Bardek}.
 The Hamiltonian (8) is not hermitian in
 terms of the collective field $\rho({\bf r})$ and $\pi({\bf r})$ due to
 the presence of two imaginary terms. In order to obtain the hermitian 
Hamiltonian, we have to rescale the wave functionals using the Jacobian 
of the transformation from ${\bf r}_{i}$ to $\rho({\bf r})$:
\begin{equation}
H\longrightarrow J^{\frac{1}{2}}HJ^{-\frac{1}{2}}\;\;.
\end{equation}
The Jacobian $J$ is determined from hermiticity condition and is given by 
\begin{equation}
\ln J=\frac{\lambda-2}{4}\int d^2 r \rho({\bf r})\ln\rho({\bf r})+
\frac{\lambda}{2}\int\int d^2 r d^2 r' 
\rho({\bf r})\ln|{\bf r}-{\bf r'}|  \rho({\bf r'})\;\;.
\end{equation}
The straightforward algebra gives the hermitian Hamiltonian  
\[
H=\frac{1}{2}\int d^2r \rho({\bf r}) (\vec{\nabla}\pi)^2+
\frac{1}{2} \int d^2r\rho({\bf r})\left[\frac{2-\lambda}{4}
\frac{\vec{\nabla}\rho({\bf r})}{\rho({\bf r})}
-\lambda\int d^2r'\frac{{\bf r}-{\bf r'}}
{|{\bf r}-{\bf r'}|^2}\rho({\bf r'})\right]^2+
\]
\begin{equation}
+\frac{\omega^2}{2}\int d^2r \rho({\bf r}){\bf r}^2+
\frac{2-\lambda}{8}\int\int d^2r d^2r' \delta({\bf r}-{\bf r'})\Delta
 \delta({\bf r}-{\bf r'})-\lambda \pi N \delta({\bf 0})\;\;.
\end{equation}
The last two singular terms do not give a contribution in the leading order
in $1/N$ expansion and should be canceled by the infinite zero-point energy of
the collective field $\rho({\bf r})$.

\section{Ground state and fluctuations}
In order to find the ground state energy of the system, we assume that the
corresponding collective field is static and has consequently a vanishing
momentum $\pi({\bf r})$. Therefore, the leading part of the collective
Hamiltonian (11) in the $1/N$ expansion is given by the effective potential
\begin{equation}
V_{\it eff}=\frac{1}{2}\int d^2r \rho({\bf r})
\left[ \frac{2-\lambda}{4} \frac{\vec{\nabla} \rho({\bf r})}{\rho({\bf r})}
-\lambda\int d^2r'
 \frac{{\bf r}-{\bf r'}}{ |{\bf r}-{\bf r'}|^2} \rho({\bf r'})\right]^2
+\frac{\omega^2}{2}\int d^2r \rho({\bf r}){\bf r}^2\;.
\end{equation}
A lower bound on the energy is obtained by performing partial
integration in order to rewrite $V_{eff}$ as  
\[
V_{\it eff}=\frac{1}{2}\int d^2r \rho({\bf r})
\left[ \frac{2-\lambda}{4} \frac{\vec{\nabla} \rho({\bf r})}{\rho({\bf r})}
-\lambda \int d^2r' \frac{{\bf r}-{\bf r'}}{ |{\bf r}-{\bf r'}|^2}
\rho({\bf r'})+\omega{\bf r}\right]^2+
\]
\begin{equation}
+\frac{N(N-1)}{2}\lambda \omega+N\omega\;\;.
\end{equation}
The first term in (13) is positive semidefinite, whereas the rest is constant.
The Bogomol'nyi bound is saturated by the positive  solution
 $\rho_{0}({\bf r})$ of the equation
\begin{equation}
\frac{2-\lambda}{4} \frac{\vec{\nabla} \rho({\bf r})}{\rho({\bf r})}
-\lambda \int d^2r' \rho({\bf r'})
 \frac{{\bf r}-{\bf r'}}{ |{\bf r}-{\bf r'}|^2}
+\omega{\bf r}=0\;\;,
\end{equation}
with the ground-state energy equal to 
\begin{equation}
E_{0}=\frac{N(N-1)}{2}\lambda \omega+N\omega\;\;.
\end{equation}
This is the exact result \cite{Calogero2}. Applying the gradient operator 
to the integrodifferential equation (14), we obtain a differential equation of
the Liouville type
\begin{equation}
\frac{2-\lambda}{4} \Delta \ln \rho({\bf r}) 
-2\pi \lambda \rho({\bf r})+2\omega=0\;\;.
\end{equation}
It is evident that the character
of the solution depends crucially on the value of the parameter $\lambda$.
For $\lambda<2$, the solution has a Gaussian fall-off at large distances.
To our knowledge, its analytic form has not been found yet.
For special value of $\lambda$, i.e. $\lambda=0$  
, the equation (16) can be solved exactly by the Gaussian.
It is obvious that in the case $\lambda \neq 0$ there
always exist an interesting uniform solution of Eq.(16) given by
\begin{equation}
\rho_0=\frac{\omega}{\lambda \pi}\;\;.
\end{equation}
Needless to say, it's existence is possible on the compact support only.
This uniform background configuration describes a
condensed state of particles. From now on, we will be primarly concerned
with analysing the spectrum of low-lying excitation around the 
configuration (17).
Performing the $1/N$ expansion of the collective field
 $\rho({\bf r})$ in the form
\begin{equation}
\rho({\bf r})=\rho_0+\eta({\bf r})\;\;,
\end{equation}
where $\rho_0$ is the aforementioned uniform ground-state 
configuration and $\eta({\bf r})$ is a small density fluctuation
around $\rho_0$, we can rewrite the collective Hamiltonian (11) up to
quadratic terms in $\pi({\bf r})$ and $\eta({\bf r})$ as 
\[
H=\frac{\rho_0}{2}\int d^2r (\vec{\nabla}\pi)^2 +\frac {\rho_0}{2}\int d^2r
\left( \frac{2-\lambda}{4} \frac{\vec{\nabla} \eta({\bf r})}{\rho_0}-\lambda 
\int d^2r'\eta({\bf r'})
 \frac{{\bf r}-{\bf r'}}{ |{\bf r}-{\bf r'}|^2}\right)^2+
\]
\begin{equation}
+E_0+\frac{2-\lambda}{8}\int \int d^2r d^2r' \delta({\bf r}-{\bf r'})\Delta
\delta({\bf r}-{\bf r'})-\lambda\pi N\delta({\bf 0})\;\;.
\end{equation}
There are no terms linear in  $\eta({\bf r})$ as we expand around the minimum
of the dominant, large-N effective potential.
Calculation of discarded terms of third order gives an interaction 
between elementary excitations of a Calogero system.
Rewriting the second term in (19) we obtain
\[
H=E_0 +\frac{\rho_0}{2}\int d^2r (\vec{\nabla}\pi)^2+
\frac{(2-\lambda)^2}{32 \rho_0}\int d^2r (\vec{\nabla} \eta)^2+
\]
\[
+\frac{(2-\lambda)\lambda \pi}{2}
\int d^2r d^2r' \eta({\bf r})\delta({\bf r}-{\bf r'}) \eta({\bf r'})-
\lambda^2 \pi \rho_0 \int d^2r d^2r' \eta({\bf r}))
\ln |{\bf r}-{\bf r'}| \eta({\bf r'})+
\]
\begin{equation}
+\frac{2-\lambda}{8}\int \int d^2r d^2r' \delta({\bf r}-{\bf r'})\Delta
\delta({\bf r}-{\bf r'})-\lambda\pi N\delta({\bf 0})\;\;.
\end{equation}
Here, we have employed the identity for the two-dimensional
Green's function:
\begin{equation}
\Delta \ln |{\bf r}-{\bf r'}|=\vec{\nabla}\frac{{\bf r}-{\bf r'}}
{|{\bf r}-{\bf r'}|}=2\pi \delta({\bf r}-{\bf r'})\;\;.
\end{equation}    
Apart from the long-range two-dimensional Coulomb repulsion ($\ln r$)
this Hamiltonian contains the effective hard core,
two-dimensional $\delta$-function interaction.
As a next step we rewrite the density fluctuation $\eta({\bf r})$ and
conjugate momentum $\pi({\bf r})$ in terms of operators
$a({\bf k})$ and $a^\dag({\bf k})$
\begin{equation}
\eta({\bf r})=\sqrt{\rho_{0}}
\int d^2k \left(e^{i{\bf k} {\bf r}}a({\bf k})+
e^{-i{\bf k} {\bf r}}a^\dag({\bf k})\right)\;\;,
\end{equation}
\begin{equation}
\pi({\bf r})=-\frac{i}{2(2\pi)^{2}\sqrt{\rho_0}}
\int d^2k \left(e^{i{\bf k}{\bf r}}a({\bf k})-
e^{-i{\bf k}{\bf r}}a^\dag({\bf k})\right)\;\;,
\end{equation}
satisfying bosonic canonical commutation relations
\begin{equation}
\left[a({\bf k}),a^\dag({\bf k}')\right]=\delta({\bf k}-{\bf k}')\;\;,
\end{equation}
\begin{equation}
\left[a({\bf k}),a({\bf k}')\right]=
\left[a^\dag({\bf k}),a^\dag({\bf k}')\right]=0\;\;.
\end{equation}
The Hamiltonian becomes
\[
H=E_0+\int d^2k \epsilon({\bf k}) a^\dag({\bf k})a({\bf k})+
\frac{1}{2}\int d^2k \Delta({\bf k})
\left(a({\bf k})a(-{\bf k})+a^\dag({\bf k})a^\dag(-{\bf k})\right)+
\]
\begin{equation}
+\delta({\bf 0})\int d^2k \epsilon({\bf k})
+\frac{2-\lambda}{8}\int \int d^2r d^2r' \delta({\bf r}-{\bf r'})\Delta
\delta({\bf r}-{\bf r'})-\lambda\pi N\delta({\bf 0})\;\;,
\end{equation}
with
\begin{equation}
\epsilon({\bf k})=\frac{{\bf k}^2}{16 \pi^2}+
\left( \frac{2-\lambda}{2}\pi|{\bf k}|+
\frac{4 \pi^{2} \lambda \rho_{0}}{|{\bf k}|} \right)^2\;\;,
\end{equation}
\begin{equation}
\Delta({\bf k})=-\frac{{\bf k}^2}{16 \pi^2}+
\left( \frac{2-\lambda}{2}\pi|{\bf k}|+
\frac{4\pi^2\lambda \rho_0}{|{\bf k}|}\right)^2\;\;.
\end{equation}
It contains non-diagonal term and  divergent normal reordering correction to
the ground state energy, proportional to $\delta({\bf 0})$, i.e. to the
area of the system. To proceed further, we have to introduce the new operators
$b({\bf k})$ and $b^\dag({\bf k})$, which are related to $a({\bf k})$ and
$a^\dag({\bf k})$ by a Bogoliubov transformation \cite{Bogoliubov}
\begin{equation}
a({\bf k})=\alpha_{\bf k} b({\bf k})+\beta_{\bf k} b^\dag(-{\bf k})\;\;,
\end{equation}
\begin{equation}
a^\dag({\bf k})=\alpha_{\bf k} b^\dag({\bf k})+\beta_{\bf k} b(-{\bf k})\;\;.
\end{equation}
These new operators satisfy cannonical commutation relations
\begin{equation}
\left[b({\bf k}),b^\dag({\bf k'})\right]=\delta({\bf k}-{\bf k'})\;\;,
\end{equation}
\begin{equation}
\left[b({\bf k}),b({\bf k'})\right]=
\left[b^\dag({\bf k}),b^\dag({\bf k'})\right]=0\;\;.
\end{equation}
provided that
\begin{equation}
\alpha_{\bf k}^2-\beta_{\bf k}^2=1\;\;.
\end{equation}
The coefficients $\alpha_{\bf k}$ and $\beta_{\bf k}$ of the Bogoliubov
 transformation are
determined by requirement that non-diagonal term of the Hamiltonian (26)
expressed in terms of operators $b({\bf k})$ and $b^\dag({\bf k})$ vanish and
this condition leads to
\begin{equation}
\alpha_{\bf k}^2=\frac{\epsilon({\bf k})+\omega({\bf k})}{2\omega({\bf k})}
\;\;,
\end{equation}
\begin{equation}
\beta_{\bf k}^2=\frac{\epsilon({\bf k})-\omega({\bf k})}{2\omega({\bf k})}\;\;.
\end{equation}
Then the Hamiltonian is diagonalized and reads
\[
H=E_0+\int d^2k \omega({\bf k}) b^\dag({\bf k})b({\bf k})+
\delta({\bf 0}) \int d^2k \frac{\omega({\bf k})}{2}+
\]
\begin{equation}
+\frac{2-\lambda}{8}\int \int d^2r d^2r' \delta({\bf r}-{\bf r'})\Delta
\delta({\bf r}-{\bf r'})-\lambda\pi N\delta({\bf 0})\;\;,
\end{equation}
where $\omega({\bf k})$ represents the energy spectrum of the physical
fluctuations
\begin{equation}
\omega({\bf k})=
\sqrt{\epsilon^2({\bf k})-\Delta^2({\bf k})}=
\left| \frac{2-\lambda}{4} {\bf k}^2 +2\omega \right|\;\;.
\end{equation}
The ground state $|0 \rangle$ is defined by
\begin{equation}
b({\bf k})|0\rangle =0 
\end{equation}
and $b^\dag({\bf k})$ is the creation operator of the low-lying excitation
with momentum ${\bf k}$.

\section{Dispersion law and divergencies}
We note that there are two qualitatively different dispersion laws,
depending on the value of the coupling $\lambda$. For $\lambda<2$, the function
$\omega({\bf k})$ has one stationary point at ${\bf k}={\bf 0}$ (absolute
minimume)
\begin{equation}
\omega({\bf k})=\frac{(2-\lambda)}{4} {\bf k}^2+2\omega ,\;
\lambda <2 \;\;.
\end{equation}
For $\lambda >2$, the function $\omega({\bf k})$ has one stationary point
at ${\bf k}={\bf 0}$ and infinite number of zero points located at the
circle with critical radius $k_{c}=\sqrt{\frac{8\omega}{\lambda-2}}$:
\begin{equation}
\omega({\bf k})=
\left\{ \begin{array}{l}
\frac{2-\lambda}{4}{\bf k}^{2}+2\omega\;\;,\;|{\bf k}|\leq k_{c}\\
\frac{\lambda-2}{4}{\bf k}^{2}-2\omega\;\;,\;|{\bf k}|>k_{c}
	\end{array}\right.,\;
\lambda>2\;\;.
\end{equation}
Thus, there is a discontinuity in slope at $k=k_{c}$.
The minimum of $\omega({\bf k})$ touches zero at $|{\bf k}|=k_c$. The
appearance of such gapless excitation with nonzero momentum can be 
considered as an indication of instability of the system to 
phase separation, or collaps. This result clearly shows the breakdown of
the Bogoliubov approximation near the critical radius $|{\bf k}|=k_c$.
The coefficients $\alpha_{\bf k}$ and $\beta_{\bf k}$ diverge there,
indicating that collective-field dispersion cannot be trusted 
for large values of the quasiparticle momentum.  
We note that in both phases there is the plasmon gap at ${\bf k}={\bf 0}$, 
generated by the induced two dimensional long-range Coulombic 
interactions.
As long as $\lambda>2$, the low-momentum excitation spectrum (40) produces
a minimum between the 'plasmon' frequency $2\omega$ and the
 'free-particle' limit $\omega \sim {\bf k}^2/2m^*$, where $m^*$ is
the effective mass of the quasi-particle:
\begin{equation}
m^*=\frac{2}{|2-\lambda|} m\;\;.
\end{equation}
Notice that the quasiparticle effective mass $m^*$ increases with 
$\lambda$ exhibiting a divergence at a critical value
$\lambda=2$, after which it decreases with increasing $\lambda$.
The above dispersion relations have interesting interpretation in terms
of the effective, Fermi-sea excitations.
A single hole-like excitation is obtained by moving a state from
the Fermi sea ($|{\bf k}|<k_c$) to the Fermi surface ($|{\bf k}|=k_c$),
while a single-particle-like excitation is obtained by moving a state from
the Fermi surface to a level above ($|{\bf k}|>k_c$).
There is nothing strange about the presence of such fermionic features in our
genuine bosonic picture. After all, we have assumed that our wavefunctions
incorporate hard-core condition in the form of the Jastrow prefactor.

Now we consider the problem of singular contributions present in the 
Hamiltonian (36).   
Using the integral representation
of the Dirac delta function we can rewrite the singular part of the 
Hamiltonian (36) as:
\[
H_{sing.}=\frac{2-\lambda}{8}\left.\left.\int d^2r \Delta \delta({\bf r})
\right|_{{\bf r}={\bf 0}}-\lambda \pi N  \delta({\bf r})
 \right|_{{\bf r}={\bf 0}}=
\]
\begin{equation}
=\frac{\lambda-2}{8}\frac{A}{(2\pi)^2}\int d^2 k {\bf k}^2-
\frac{\lambda \pi N}{(2\pi)^2}\int d^2k\;\;,
\end{equation}
where $A$ denotes the infinite area of the system
\begin{equation}
A=\int d^2r=(2\pi)^2 \left. \delta({\bf k})\right|_{{\bf k}={\bf 0}}\;\;.
\end{equation}
Since the total number of particles $N$ and the area $A$ are interrelated
by relation (17)
\[
\rho_{0}=\frac{N}{A}=\frac{\omega}{\lambda \pi}\;\; ,
\]
the singular part (39) reduces to
\begin{equation}
H_{sing.}=\left. \frac{\delta({\bf k})}{2} \right|_{{\bf k}={\bf 0}}
\int d^2k \left(\frac{\lambda-2}{4}{\bf k}^2-2\omega\right)\;\;.
\end{equation}
In the case of the first phase $\lambda<2$, it is evident that the singular
contribution $H_{sing.}$ completely cancels the divergent normal
reordering correction. There is no correction to the ground state energy due to
quadratic fluctuations.
Naively, in the phase $\lambda>2$ there is no complete
cancelation and the correction to the ground state energy is found to be:
\begin{equation}
\Delta E_{0}=\left. \delta({\bf k}) \right|_{{\bf k}={\bf 0}}
 \frac{\lambda-2}{4}\int_{|{\bf k}|=k_c}^{\infty}
d^2k ({\bf k}^2-k_c^2)\;\;. 
\end{equation}
However, taking into account the relation (17) and the definition of $k_c$
we see that in our approach the sector $ |{\bf k}|>k_c$ is absent because of
large-N limit, i.e. $k_c^2\sim \rho_0\rightarrow \infty$.
There is no correction to the ground state energy again.  
The finite Hamiltonian of quantum collective excitations reduces to
\begin{equation}
H=\frac{N(N-1)}{2}\lambda \omega+N \omega +
 \int d^2k \frac{2-\lambda}{4} {\bf k}^2 b^\dag ({\bf k}) b({\bf k})+
2\omega {\cal N}\;\;,
\end{equation}
where the number operator ${\cal N}$ labels the quanta of
oscillator excitations:
\begin{equation}
{\cal N}=\int d^2k {b}^{\dag}({\bf k}) b({\bf k})\;\;.
\end{equation}
The excitation spectrum of density fluctuations around the
semiclassical solution $\rho_0$ shows roughly the infinite tower structure of
eigenvalues, separated by the spacing $2\omega$ \cite{Calogero2,Ghosh1}.
Each level is smeared by small, continuous contributions from
kinetic energies of particles ($\lambda<2$) or holes ($\lambda>2$).

\section{Wavefunctionals}
Now we are in a position to find the collective-field vacuum wave functional.
With the definition (38) and the Bogoliubov transformation (29) and (30),
we get 
\begin{equation}
b({\bf k})|0 \rangle=(\alpha_{\bf k} a({\bf k})-\beta_{\bf k}
 a^{\dag}(-{\bf k}))
|0 \rangle=0
\;\;.
\end{equation}
Rewriting the bosonic operators $a({\bf k})$ in terms of the density
fluctuation $\eta({\bf r})$ and conjugate momentum $\pi({\bf r})$ the vacuum
condition (48) becomes 
\begin{equation}
\int d^2r e^{i{\bf kr}}\left[ (\alpha_{\bf k}-\beta_{\bf k})\eta({\bf r})+
2 (2 \pi)^2 \rho_0 (\alpha_{\bf k}+\beta_{\bf k})
\frac{\delta}{\delta \eta({\bf r})}\right] |0\rangle =0\;\;.
\end{equation}
It is evident that solution of this linear functional derivative equation
is given by the Gaussian ansatz:
\begin{equation}
|0 \rangle=e^{ \int d^2 r d^2 r' \eta({\bf r}) K({\bf r}-{\bf r'}) 
\eta({\bf r'}) }
\end{equation}
with translationally invariant kernel $K({\bf r})$. Using Fourier transform 
\begin{equation}
K({\bf r}-{\bf r'})=\int d^2 k e^{i{\bf k}({\bf r}-{\bf r'})}\tilde{K}({\bf k})
\end{equation}
and combining relations (49), (50) and (51), we get
\begin{equation}
\int d^2 r e^{i{\bf kr}} \eta({\bf r})
\left[ (\alpha_{\bf k}-\beta_{\bf k})+
4\rho_0 (2 \pi)^4 (\alpha_{\bf k}+\beta_{\bf k})\tilde{K}({\bf k})
\right]=0\;\;\;.    
\end{equation}
Since this equation holds for any ${\bf k}$, the kernel $\tilde{K}({\bf k})$
is given by
\begin{equation}
\tilde{K}({\bf k})=-\frac{1}{4(2\pi)^4 \rho_0}
\frac{\alpha_{\bf k}-\beta_{\bf k}}{\alpha_{\bf k}+\beta_{\bf k}}=
-\frac{1}{2(2 \pi)^2 \rho_0}\left|\frac{\lambda-2}{4}-\frac{2\omega}{k^2}
\right|\;\;. 
\end{equation}
This result enables us to reconstruct the Schr\"{o}dinger wave function 
$\psi({\bf r}_1,{\bf r}_2,...,{\bf r}_N)$ for the ground-state of the 
N-particle system. The ground state functional is given by the
\begin{equation}
\Psi[\rho]=\prod_{i>j}|{\bf r}_i-{\bf r}_j|^\lambda J^{-\frac{1}{2}}|0\rangle
\;\;.
\end{equation}   
Here, the prefactor is present owing to the extraction (2). The
Jacobian of the transformation from ${\bf r}_i$ to $\rho({\bf r})$ rescales
wavefunctional by the factor $J^{\frac{1}{2}}$. Expanding the Jacobian to
the quadratic term in $\eta$ and using relations (10), (17) and the Bogomol'ny
equation for $\rho_0$ (14), we are left with
\[
\Psi[\eta]=\prod_{i>j}|{\bf r}_i-{\bf r}_j|^\lambda 
\exp
 \left[ -\frac{\omega}{2} \int d^2 r {\bf r}^2 \eta({\bf r})-\frac{\lambda}
{2}\int \int d^2 r d^2 r' \eta({\bf r})\ln|{\bf r}-{\bf r'}|
\eta ({\bf r'})+\right.
\]
\begin{equation}
\left.
+\frac{2-\lambda}{8 \rho_0}\int d^2r \eta^2({\bf r})+
\int\int d^2r d^2r' \eta({\bf r}) K({\bf r}-{\bf r'})
\eta({\bf r'}) \right]
\end{equation}         
Next, we employ the Fourier representation of the $\ln$ 
kernel:
\begin{equation}
\ln|{\bf r}-{\bf r'}|=-\frac{1}{2 \pi}\int d^2k 
\frac{e^{i{\bf k}({\bf r}-{\bf r'})}}{k^2}\;\;,
\end{equation}
so that the vacuum functional $\Psi$ can be rewritten in the $\lambda <2$
phase as 
\begin{equation}
\Psi=\prod_{i>j}|{\bf r}_i -{\bf r}_j |^\lambda 
e^{-\frac{\omega}{2}\int d^2r {\bf r}^2 \eta({\bf r})}\;\;.
\end{equation}
If we substitute equation (4) in (57), we obtain the well known ground-state
wavefunction for N-particles \cite{Calogero2}:
\begin{equation}
\Psi({\bf r}_1,{\bf r}_2,...,{\bf r}_N)=
\prod_{i>j}|{\bf r}_i -{\bf r}_j |^\lambda e^{-\frac{\omega}{2}\sum_i
 {\bf r}_i^2}\;\;.
\end{equation}
In order to find the vacuum wave-functional in the  $\lambda>2$ phase, we
rearrange the contributions from integrations in (55) and obtain the following
expression
\[
\Psi[\eta]=\prod_{i>j}|{\bf r}_i -{\bf r}_j |^\lambda 
\exp \left(-\frac{\omega}{2} \int d^2r \eta({\bf r}){\bf r}^2\right)
\times
\]
\begin{equation}
\times
\exp \left[
\frac{\lambda}{2\pi} \int\int d^2r d^2r' \eta({\bf r}) \eta({\bf r'})
\left( \int_{|{\bf k}>k_c}d^2k 
\frac{e^{i{\bf k}({\bf r}-{\bf r'})}}{|{\bf k}|^2}-    
\int_{|{\bf k}|>k_c}
d^2k \frac{e^{i{\bf k}({\bf r}-{\bf r'})}}{k_c^2} \right)
\right]
\;\;.
\end{equation}
We get the same vacuum wave-functional as in the $\lambda<2$ phase (57),
since $k_c\rightarrow \infty$ in the large-N limit.
The unrenormalized wave function of the first excited state may be 
represented by
\begin{equation}
\psi({\bf k})=b^\dag ({\bf k})|0 \rangle\;\;.
\end{equation}
Using the complex conjugate of relation (48) it is easily shown that
the phonon state is mearly the coordinate multiplied by the Gaussian form:
\begin{equation}
 \psi({\bf k})=\frac{2}{\sqrt{\rho_0}}(\alpha _{\bf k}-\beta_{\bf k})
\int d^2r \eta({\bf r}) e^{i{\bf kr}} |0 \rangle\;\;.
\end{equation}
When expressed in configuration space coordinates, these excited states
are just the functions used by Feynman to describe phonon and roton states
\cite{Feynman}:
\begin{equation}
 \psi({\bf k})\sim \sum_i e^{i{\bf k}{\bf r}_i} |0 \rangle\;\;.
\end{equation}
We emphasize here that the one-particle excited states have
vanishing norm at $|{\bf k}|=k_c$.
This is the consequence of the fact that $|{\bf k}|=k_c$ is
the singular point of the Bogoliubov transformation (34) and 
(35) ($\alpha(k_c)=\beta(k_c)=\infty$).
However, these large momentum excitations are beyond the scope of our
approximation.

\section{Correlation function and static structure factor}
Next, we study the density-density correlation function and the static
structure factor in the 2D CM model. The analysis is similar to the 
one dimensional case \cite{Andric}. The density$-$density correlation function
$G({\bf r})$ is defined by
\begin{equation}
\langle 0|\rho({\bf r})\rho({\bf r'})|0 \rangle=
\rho_0^2+ \langle 0|\eta({\bf r}) \eta({\bf r'}) | 0 \rangle=
\rho_0^2+G({\bf r}-{\bf r'})\;\;.
\end{equation}
The relations (22), (29) and (30) enable us to evaluate the function
$G({\bf r})$ 
\begin{equation}
G({\bf r})=\frac{\rho_0}{(2\pi)^2} \int d^2k e^{i{\bf kr}}
\frac{{\bf k}^2}{2 \omega({\bf k})}\;\;.
\end{equation}  
The static structure factor $S({\bf k})$ is defined as the 
Fourier transform of $G({\bf r})$:
\begin{equation}
S({\bf k})=\rho_0 \frac {{\bf k}^2}{2\omega({\bf k})}\;\;.
\end{equation}
This is the familiar Feynman-Bijl relation.
By expanding the static structure factor $S({\bf k})$ in powers
of ${\bf k}^2/\rho_0$ up to the cubic terms, we easily obtain
\begin{equation}
S({\bf k})=\frac{{\bf k}^2}{4 \lambda \pi}
\left( 1\mp \frac{(2-\lambda)}{8\lambda \pi \rho_0} {\bf k}^2
+\frac{(2-\lambda)^2}{(8\lambda \pi \rho_0)^2} {\bf k}^4\mp...\right)
\end{equation}
where the upper and lower sign refers to the $\lambda<2$ and 
$\lambda>2$ phase, respectively.
We are now in position to compare the collective field theory structure
factor with the exact one, known from the random matrix theory for 
special value  $\lambda=1$ \cite{Khare1,Mehta,Ginibre}. Namely, it is
 easily seen, that the squared value of the ground-state wavefunction (58)
 is proportional to the joint probability density function of the
 eigenvalues of  complex matrices, provided one sets 
$\lambda$ equal to $1$. By expanding the exact static structure factor
$S_{\lambda=1}({\bf k})$:
\begin{equation}
S_{\lambda=1}({\bf k})=
\rho_0 \left(1-e^{-\frac{{\bf k}^2}{4\pi \rho_0}} \right)\;\;,
\end{equation}
we obtain
\begin{equation}
S_{\lambda=1}({\bf k})=\rho_0 \left(
\frac{{\bf k}^2}{4 \pi\rho_0}-\frac{{\bf k}^4}{2(4 \pi \rho_0)^2}+
\frac{{\bf k}^6}{6(4\pi \rho_0)^3}-...\right)\;\;.
\end{equation}
Comparing relations (66) and (68), one immediately sees that they agree
up to forth order for $|{\bf k}|$ close to $0$. 
This indicates that our method is meaningful as long 
as we restrict ourselves to the wavelengths of density fluctuations
much longer than the average inter-particle separation $1/\sqrt{\rho_0}$.
Having found the static structure factor, we now turn to the 
calculation of the correlation function $G({\bf r})$.
Using the definition (64) and expansion (66) we find that 
\begin{equation}
G({\bf r})=-\frac{1}{(2\pi)^2}\int d^2k e^{i{\bf kr}}
\left[\frac{1}{4 \lambda \pi} {\bf k}^2 \mp
\frac{(2-\lambda)}{32 \lambda^2 \pi^2 \rho_0} {\bf k}^4+...\right]\;\;,
\end{equation}
The asymptotic form of $G({\bf r})$ at large values of $|{\bf r}|$
gets only a contribution from the small $|{\bf k}|$ region in the integral
(69). Using the definition of the Bessel functions of the first kind
of order zero, it can be seen that the correlation function shows long-range
order at large distances for any value of $\lambda \neq 1$:
\begin{equation}
G({\bf r})\sim -\frac{1}{8 \lambda \pi^2 r^4}
\int_0^1 dx x^3 J_0(x)\pm \frac{(2-\lambda)}{64 \lambda^2 \pi^3 \rho_0 r^6}
\int_0^1 dx x^5 J_0(x)+...
\end{equation}

\section{Conclusion}
In this paper we have formulated a quantum field theory of two-dimensional
CM model in terms of collective-field variables in the large-N limit.
In this semiclassical approach, we have determined the ground-state
energy and the corresponding uniform density solution.
Using the method of the Bogoliubov approximation, we have found the
spectrum of low-lying excitations around the ground state for
any value of the coupling constant $\lambda$. We have shown that the
spectrum splits into two branches depending on the sign of $\lambda-2$.
We have proposed the interpretation of the spectrum in terms of quasi-particle
($\lambda<2$) and quasi-hole ($\lambda>2$) excitations in the effective
Fermi sea, defined by the Fermi momentum $k_c$. We have also shown
that the ground-state energy is not affected by the next to leading
order corrections steaming from the quantum collective-field
fluctuations. 
Furthermore, we have obtained the density-density correlation
function and the static structure factor which shows long-range
order.  
Finally, we believe that the exact $\omega({\bf k})$
does not vanish at nonzero value of $|{\bf k}|$ for $\lambda>2$.
The correct dispersion probably has a roton-like minimum at 
$|{\bf k}|=k_c$. In order to test the existence of roton dip
in the spectrum of the excitation one has to go beyond the
quadratic approximation. We hope to report on this issues in a
separate publication.

\section*{ Acknowledgment}
We would like to thank Larisa Jonke for fruitful discussions on
the subject.
This work was supported by the Ministry of Science of Republic
of Croatia under contract No. 0098003.

\end{document}